\documentstyle[12pt]{article}
\newcommand{\be}{\begin{equation}}
\newcommand{\bel}[1]{\begin{equation}\label{#1}}
\newcommand{\ee}{\end{equation}}
\newcommand{\bea}{\begin{eqnarray}}
\newcommand{\ba}{\begin{array}}
\newcommand{\eea}{\end{eqnarray}}
\newcommand{\ea}{\end{array}}

\newcommand{\BEQ}{\begin{equation}}    
\newcommand{\BEA}{\begin{eqnarray}}
\newcommand{\EEQ}{\end{equation}}      
\newcommand{\EEA}{\end{eqnarray}}
\newcommand{\sig}{\sigma}                        
\newcommand{\rar}{\rightarrow}                   
\newcommand{\wit}[1]{\widetilde{#1}}             
\newcommand{\ket}[1]{\left|#1\right\rangle}      
\newcommand{\bra}[1]{\left\langle #1\right|}     
\newcommand{\zeile}[1]{\vskip #1 \baselineskip}  
\newcommand{\vekz}[2]
     {\mbox{${\begin{array}{c} #1  \\ #2 \end{array}}$}}
\newcommand{\matz}[4]
     {\mbox{${\begin{array}{cc} #1 & #2  \\ #3 & #4 \end{array}}$}}

\newcommand{\build}[3]{\mathrel{\mathop{\kern 0pt#1}\limits_{#2}^{#3} }}

\newcommand{\appsection}{\setcounter{equation}{0} \section*{Appendix}
\renewcommand{\theequation}{A.\arabic{equation}}
              \renewcommand{\thesection}{A} }
\catcode`\@=11
\def\numberbysection{\@addtoreset{equation}{section}
        \def\theequation{\thesection.\arabic{equation}}}

\begin{document}
\baselineskip 0.3in
%
%
\begin{titlepage}
\null
\vskip 1cm
\begin{center}
\vskip 0.5in
{\Large \bf Equivalences between stochastic systems}
\vskip 0.5in
Malte Henkel, Enzo Orlandini and Gunter M. Sch\"utz
 \\[.3in]
{\em Theoretical Physics, Department of Physics, \\
University of Oxford, 1 Keble Road, Oxford OX1 3NP, UK}
\end{center}
\zeile{2}
%
\begin{abstract}
Time-dependent correlation functions of (unstable) particles undergoing
biased or unbiased diffusion, coagulation and annihilation are calculated.
This is achieved by  similarity transformations between different
stochastic models and between stochastic and soluble
{\em non-stochastic} models.
The results agree with experiments on one-dimensional
annihilation-coagulation processes.
\end{abstract}
\zeile{3}
\end{titlepage}

\newpage
%
%
The physics of interacting particles out of thermodynamic equilibrium
has been a subject of much recent interest. While in larger spatial
dimensions, conventional rate equation approaches are sufficient, systems
constrained to be effectively one-dimensional display novel and interesting
fluctuation effects. For example, for particles $A$ diffusing on a lattice
and undergoing a binary reaction process $A+A\rar\mbox{\rm products}$
one expects for
large times $t$ an algebraic fall-off of the mean particle concentration
\BEQ \label{LarT}
\bar{c}(t) \sim t^{-y}
\EEQ
In $1D$ systems, one finds $y=\frac{1}{2}$ as opposed to $y=1$ obtained
from a (mean-field) rate equation. The exponent $y$ has also
been measured experimentally in effectively one-dimensional systems.
For annihilation-coagulation
reactions $A+A\rar \mbox{\rm products}$
one finds $y=0.52-0.59$ \cite{Pras89} and $y=0.47(3)$ \cite{Kopel90},
and for the (pure) coagulation reaction
$A+A\rar A$, $y\simeq 0.48$ \cite{Kroon93}.
Although these reacting systems might appear to be quite different, in this
work we show that these and more general systems can be treated in a
simple and unified way. In particular, a simple explanation for the same
value of $y$ in all annihilation-coagulation problems is obtained.

For the theoretical description of these reaction-diffusion systems,
a useful approach \cite{QHF}
consists of rewriting the master equation which
describes the time evolution of the probability distribution function
$P(\{\beta\};t)$ as a Schr\"odinger equation
\BEQ\label{ME}
\partial_t P(\{\beta\};t) = - H P(\{\beta\};t)
\EEQ
in which the quantum Hamiltonian $H$ is defined in terms of the
transition rates $w(\beta \rar \gamma)$ between two configurations
$\{\beta\}$ and $\{\gamma\}$ by
\BEQ
\bra{\gamma} H \ket{\beta} = - w(\beta\rar\gamma) \;\; , \;\;
\bra{\beta} H \ket{\beta} = \sum_{\gamma\neq\beta} w(\beta\rar\gamma) \; \; .
\EEQ
$H$ describes a stochastic process since
the sums of all elements in each column add up to zero.
This conservation of probability is equivalent to the relation
\BEQ \label{LeftGS}
\bra{s} H = 0 \;\; , \;\; \bra{s} = \sum_{\beta} \bra{\{\beta\}}
\EEQ
for the vector $\bra{s}$. Then the well-known
theorems about the solutions of the master equation \cite{Hyver} can be
translated into the Hamiltonian formulation at hand. In particular,
the real parts of the eigenvalues of $H$ are non-negative.
Furthermore, starting from an initial probability distribution
$\ket{P_0}=\sum_{\beta}
P(\{\beta\};t=0) \ket{\{\beta\}}$ where each configuration $\{\beta\}$
occurs with probability $P(\{\beta\};t=0)$, the solution to the
master equation (\ref{ME}) is then formally given by
the time-dependent probability distribution (state vector)
$\ket{P}=\sum_{\beta} P(\{\beta\};t) \ket{\{\beta\}}=\exp{(-Ht)}\ket{P_0}$.
Time-dependent averages of an
observable\footnote{$F$ is a suitably chosen
projection operator (see below).}
 $F$ are given by the matrix element
\BEQ
<F>(t) = \bra{s} F e^{-H t} \ket{P_0}
\EEQ

The interest in this setup of the problem in $1D$ comes from the
integrability of the quantum Hamiltonian $H$ for large classes of
reaction-diffusion processes \cite{Alca94,Pesch,Schue95}.
Exactly known results such as the knowledge
of the spectrum of $H$ (obtained e.g. from the Bethe ansatz)
have so far led to a number of exact and
explicit results for time-dependent averages and
correlations.
Here we show how these and other existing results (see e.g.
\cite{alt,Schue95a} for annihilation and
coagulation processes) can
be extended to considerably more general quantum Hamiltonians:
\begin{enumerate}
\item We investigate relations between stochastic systems whose quantum
Hamiltonians $H$ and $\wit{H}$ are related through a change of basis
of the one-particle states, see \cite{Alca94,Pesch,Schue95,Krebs95,Simon95}
\BEQ \label{BTra}
\wit{H} = {\cal B} H {\cal B}^{-1} \;\; , \;\;
{\cal B} = \bigotimes_{i=1}^{L} B_i
\EEQ
where $B_i$ is acting only on the site $i$.
\item We consider a given quantum (and in general non-stochastic)
Hamiltonian with known properties
and we look for
stochastic processes which can be obtained from this Hamiltonian by a
similarity transformation of the form (\ref{BTra}). The time-dependent
behaviour of these new stochastic systems can then be elucidated in terms
of the original Hamiltonian.
\end{enumerate}
Under the similarity transformation $\cal B$ from (\ref{BTra})
averages transform as follows
\BEQ
<F>(t) = \bra{s} F e^{-Ht}\ket{P_0} =
\bra{s} \wit{F} e^{-\wit{H}t} \ket{\wit{P_0}}
\EEQ
with the transformed observable $\wit{F} = F {\cal B}^{-1}$ and transformed
initial distribution $\ket{\wit{P_0}} = {\cal B}\ket{P_0}$.

We now give the general form of the single-site matrix $B$
for transformations between two stochastic systems. Certainly,
one-site state vectors of the system $S$ described by the Hamiltonian $H$
must have the form $\ket{\rho}= \left(\vekz{1-\rho}{\rho}\right)$,
with $0\leq \rho\leq1$. Also, for the transformed state $B\ket{\rho}$,
probabilities must sum up to one, for all values of $\rho$, thus
\BEQ
b_{11}(1-\rho) + b_{21}\rho + b_{12}(1-\rho) + b_{22}\rho =1
\EEQ
where the $b_{ij}$ are the elements of $B$.
Comparing coefficients, we get for $B$ the form
\BEQ
B = \left( \matz{b_1}{1-b_2}{1-b_1}{b_2} \right) \;\; .
\EEQ
Since obviously $\bra{s}B=\bra{s}$, it follows that $\bra{s}\wit{H}=0$
for the full system defined on $L$ sites.

The stochastic systems $S=(H,\rho)$ under consideration are described
by the quantum Hamiltonian $H$ and a set of parameters $\rho$ specifying the
initial conditions (see below).
The relations between two stochastic systems $S$ and $\wit{S}$
are caught by the following two definitions.
\begin{enumerate}
\item The transformation $S \rar \wit{S}$ between
two stochastic systems $S$ and $\wit{S}$
is called a {\em similarity transformation}, if
there exists a non-singular transformation
${\cal B}$ of the form (\ref{BTra})
between the quantum Hamiltonians $H$ and $\wit{H}$
such that all reaction-diffusion rates are positive in
both systems. $S$ and $\wit{S}$ are then called {\em similar}.
\item The transformation $S\rar\wit{S}$ between two systems $S$ and $\wit{S}$
is called a {\em stochastic similarity transformation (SST)}, if $S$
and $\wit{S}$
are similar and furthermore if for all
probability distributions $\ket{P}$ of $S$ also
$\ket{\wit{P}} = {\cal B}\ket{P}$ is a probability distribution of $\wit{S}$.
\end{enumerate}
Through similarities and SST a given system may be mapped into a simpler form.
Examples will be given below.
To illustrate the second definition, consider an uncorrelated initial state
of the form
\BEQ \label{IniS}
\ket{P_0} = \bigotimes_{i=1}^{L} \left( \vekz{1-\rho_i}{\rho_i} \right)
\EEQ
with $0\leq \rho_i \leq 1$ for all sites $i=1,\ldots,L$. If $\wit{S}$ is
obtained from $S$ by a SST,
initial states of the form (\ref{IniS}) are mapped
onto transformed initial states $\ket{\wit{P_0}}$ of the same form and
with $0\leq \wit{\rho}_i \leq 1$ for all sites.
We stress that the notion of a SST between two stochastic systems
$S$ and $\wit{S}$ is considerably more restrictive than mere
similarity, which does not require that also the transformed state
vector is a probability distribution.
We also remark that because of the locality of the change of basis
the results obtained here are valid in an arbitrary number of space
dimensions, although we shall present the argument explicitly only for
$d=1$.

We now state precisely the models we shall study below. Consider a
one-dimensional lattice, with $L$ sites and periodic boundary conditions.
Each lattice site can either
be empty (denoted by $\emptyset$) or occupied by a single particle
(denoted by $A$). Particles can hop to an empty nearest neighbor site.
A single particle or a pair of particles on neighboring sites can
undergo a chemical reaction. The reactions we are going to consider
are specified with their rates in table~\ref{tab1}.

\begin{table}
\begin{center}
\begin{tabular}{|l|c|c|} \hline
diffusion to the left & $\emptyset + A \rar A +\emptyset$ & $D_L$ \\
diffusion to the right& $A+\emptyset \rar \emptyset + A$  & $D_R$ \\
pair annihilation     & $A+A\rar \emptyset + \emptyset$   & $2\alpha$ \\
coagulation to the right& $A+A\rar\emptyset + A$          & $\gamma_R$ \\
coagulation to the left & $A+A\rar A+\emptyset $          & $\gamma_L$ \\
death                   & $A+\emptyset\rar\emptyset +\emptyset$& $\delta$ \\
                        & $\emptyset+A\rar\emptyset +\emptyset$& $\delta$ \\
decoagulation to the left& $\emptyset + A\rar A+A $       & $\beta_L$ \\
decoagulation to the right& $A+\emptyset\rar A+A  $       & $\beta_R$ \\
birth                   & $\emptyset+\emptyset\rar A+A$   & $2\nu$ \\
creation                & $\emptyset+\emptyset\rar\emptyset +A$ & $\sigma$ \\
                        & $\emptyset+\emptyset\rar A+\emptyset$ & $\sigma$ \\
\hline
\end{tabular}
\caption{Two-sites reaction-diffusion processes and their rates. \label{tab1}}
\end{center} \end{table}

Furthermore, we shall distinguish between unbiased and biased reactions.
For unbiased reactions, the rates with indices $L$ and $R$ are all equal
and we shall then drop the index
(e.g. $D_L= D_R=D$ etc.). For biased reactions, we define an anisotropy
parameter $\eta$ from
\BEA
D_L = D( 1+\eta) \;\; &,& \;\; \gamma_L = \gamma(1+\eta)
\;\; , \;\; \beta_L = \beta (1+\eta) \nonumber \\
D_R = D( 1-\eta) \;\; &,& \;\; \gamma_R = \gamma(1-\eta)
\;\; , \;\; \beta_R = \beta (1-\eta) \label{Kopp}
\EEA
For $\eta=0$ we recover the unbiased case.

We begin with unbiased systems, that is $\eta=0$. We consider the
following system, with diffusion, annihilation, coagulation and death
reactions present (see table~\ref{tab1}).
In what follows, we take units of time such that $D=1$. Then the Hamiltonian
can be written in terms of two-site contributions $H=\sum_i H_{i,i+1}$,
where
\BEQ \label{HMat}
H_{i,i+1} = \left( \begin{array}{cccc}
0 &  -\delta &  -\delta & -2\alpha \\
0 &  1+\delta & -1 & -\gamma \\
0 & -1 &  1+\delta & -\gamma \\
0 &  0 &  0 & 2(\alpha+\gamma) \end{array} \right)
\EEQ
We define the $k-$point correlation function
of the particle number operator $n_{x_i}$, $i=1,2,\ldots,k$ as
\BEQ
C_k(\{x\},t;\alpha,\gamma,\delta;\rho) = < n_{x_1} \cdots n_{x_k} >_{H}(t)
\EEQ
where we explicitly indicate the dependence on the rates as well as on the
initial conditions. The operator $n_x$ is a projector with eigenvalue 1
if site $x$ is occupied and eigenvalue 0 if it is vacant.
Although $H$ is non-hermitian, it is known that there is a decomposition
$H=H_{XXZ} + H_1$ into an hermitian Hamiltonian $H_{XXZ}$ (which is the
Hamiltonian of the anisotropic Heisenberg quantum spin model)
and a non-hermitian
part $H_1$ such that the eigenvalues of $H$ are exactly the eigenvalues
of $H_{XXZ}$ \cite{Alca94}.
That is so because the chemical reactions permitted here only
destroy and never create particles. In one dimension,
an interesting special case is given by the {\em free-fermion} condition
\BEQ \label{FF}
\alpha + \gamma = 1 +\delta .
\EEQ
In that case the hermitian part $H_{XXZ}$
can be diagonalized exactly in terms
of free fermions. If either just annihilation or coagulation are present,
it is known that a closed system of equations of
motion can be found \cite{alt}.
(\ref{FF}) means that diffusion and death together occur
at the same rate as annihilation and coagulation together.
If $\delta \leq \gamma$, we can rewrite the problem as an
annihilation-coagulation problem of an unstable particle, where the
effective coagulation rate is modified into
$\gamma_{\rm eff}=\gamma-\delta$, and
$1/(2\delta)$ is the life time of the unstable particle.\footnote{The special
case $\gamma=\delta$, i.e. pair annihilation of unstable particles is discussed
in the appendix.}
 If we use the
diffusion process to
determine the time scale, we can say
that if two particles attempt at the same
time to be on the same site, they undergo a chemical
reaction with probability one.
The ratio $\gamma_{\rm eff}/\alpha$ then gives the
branching ratio between the two processes.

At first sight, the condition (\ref{FF}) appears to be rather artificial.
However, it is apparently realised to good approximation in one of the
experimental realizations of the model considered so far \cite{Kroon93}.
The carrier substance is $N(CH_3)_4 MnCl_3$. The particles are excitons
of the $Mn^{2+}$ ion and move along the widely separated $MnCl_3$ chains.
A single exciton has a decay time of about
$0.7 ms$. The on-chain hopping rate is $10^{11}-10^{12} s^{-1}$.
If two excitons arrive on the same $Mn^{2+}$ ion, they undergo a
coagulation reaction with a reaction time
$\approx 100fs$ \cite{Kroon93}. Since the reaction time is much smaller
than the diffusion time, we can conclude that the reaction probability
is very close to one. This gives back (\ref{FF}), with $\alpha=\delta=0$
for this example.

After these preparatory remarks
we return to the general case. Starting from the
system $S$ as defined through its Hamiltonian (\ref{HMat}),
we get the following simplified systems $\wit{S}$ through a SST.
\begin{enumerate}
\item[I.] $\delta > 2\alpha +\gamma$. Through a SST we
get the system $\wit{S}_{I}$ with
\BEQ
\wit{D} = D \;\; , \;\;
\wit{\alpha} = 0 \;\; , \;\;
\wit{\gamma} = \alpha + \gamma \;\; , \;\;
\wit{\delta} = \delta \;\; , \;\;
\wit{\rho} = \frac{\delta-2\alpha -\gamma}{\delta-\alpha-\gamma} \,\rho
\EEQ
and we find
\BEQ
C_k(\{x\},t;\alpha,\gamma,\delta;\rho) =
\left( \frac{\delta-\alpha-\gamma}{\delta-2\alpha-\gamma} \right)^k
C_k(\{x\},t;0,\alpha+\gamma,\delta;\frac{\delta-2\alpha-\gamma}{\delta-\alpha
-\gamma} \,\rho)
\EEQ
\item[II.] $\delta = 2\alpha +\gamma$.
In this case, the transformation becomes
singular. However, the equations of motion
for the particle number correlators $C(\{x\})$ decouple from each other
\cite{Schue95}. For
example, the particle density at time $t$ is
\BEQ
C_1(x,t) = e^{-2(1+\delta)t} \sum_{m=-\infty}^{\infty}
\sum_{y=1}^L C_1(y,0) I_{x-y+m L}(2t)
\EEQ
where $I_n$ is a modified Bessel function.
\item[III.] $\delta < 2\alpha +\gamma$.
We find the transformed system $\wit{S}_{III}$
\BEQ
\wit{D} = D \;\; , \;\;
\wit{\alpha} = \alpha+\gamma \;\; , \;\;
\wit{\gamma} = 0 \;\; , \;\;
\wit{\delta} = \delta \;\; , \;\;
\wit{\rho} = \frac{2\alpha+\gamma-\delta}{2\alpha+2\gamma-\delta} \,\rho
\EEQ
and we find
\BEQ
C_k(\{x\},t;\alpha,\gamma,\delta;\rho) =
\left( \frac{2\alpha+2\gamma-\delta}{2\alpha+\gamma-\delta} \right)^k
C_k(\{x\},t;\alpha+\gamma,0,\delta;\frac{2\alpha+\gamma-\delta}{2\alpha
+2\gamma-\delta} \,\rho)
\EEQ
\item[IV.] $\delta < \alpha +\gamma$. In this interval, there is a second
SST onto the system $\wit{S}_{IV}$ with
\BEQ
\wit{D} = D \;\; , \;\;
\wit{\alpha} = \alpha+\gamma-\delta \;\; , \;\;
\wit{\gamma} = \delta \;\; , \;\;
\wit{\delta} = \delta \;\; , \;\;
\wit{\rho} = \frac{2\alpha+\gamma-\delta}{2\alpha+2\gamma-2\delta} \,\rho
\EEQ
and we find
\BEQ \label{Aeff}
C_k(\{x\},t;\alpha,\gamma,\delta;\rho) =
\left( \frac{2\alpha+2\gamma-2\delta}{2\alpha+\gamma-\delta} \right)^k
C_k(\{x\},t;\alpha+\gamma-\delta,\delta,\delta;
\frac{2\alpha+\gamma-\delta}{2(\alpha+\gamma-\delta)} \, \rho )
\EEQ
\end{enumerate}
Although both systems
$\wit{S}_{III}, \wit{S}_{IV}$ are found from $S$ through
a SST, the latter is more useful for practical
calculations. Using the results derived in the appendix,
we can isolate the dependence on
$\delta$ completely. For translationally invariant initial distributions
we have for the large-time behaviour in the free fermion case
(see the appendix for more general cases)
\BEA
C_1(x,t) &\simeq& \widetilde{\rho}_0 e^{-2 \delta t} \\
C_2(x,x+r,t) &\simeq& t^{-3/2} e^{-4 \delta t} \;\; ; \;\; r^2 \ll t \nonumber
\EEA
and we explicitly see that for $\delta \neq 0$ the initial particle density
does enter into the large-time behaviour.

This is different from the result found when the death reaction is absent
($\delta=0$). In that case only the SST onto $\wit{S}_{III}=\wit{S}_{IV}$
remains. (The corresponding similarity transformations have been
derived before \cite{Krebs95,Simon95}.)
All correlations then depend non-trivially only on the effective
annihilation rate $\alpha_{\rm eff}=\alpha+\gamma$.
The correlation function $C$ on the r.h.s. of (\ref{Aeff}) then is the known
density correlation function
for diffusion annihilation, see \cite{Alca94,alt,Schue95a}. For example, with
the initial state (\ref{IniS}) with $\rho_i = \rho$ for all sites $i$
we have for the mean particle concentration $\bar{c}(t)\sim\int dx C_1(x,t)$
in the free fermion case for the process $A+A\rar\emptyset$
\BEQ \label{AA0}
\bar{c}(t) = \rho e^{-4Dt} \left[ I_0(4Dt) +2(1-\rho) \sum_{k=1}^{\infty}
(1-2\rho)^{k-1} I_{k}(4Dt) \right] \simeq
\left( 8\pi Dt\right)^{-1/2} \left( 1+ {\cal O}(t^{-1})\right)
\EEQ
In particular,
we always get back $y=\frac{1}{2}$ in (\ref{LarT}),
in agreement with experiment \cite{Pras89,Kopel90,Kroon93}. Furthermore,
the data of Kroon et {\it al.} \cite{Kroon93}
show that the long-time behaviour of
$\bar{c}(t)$ is {\em independent} of the
initial particle density $\rho$, in agreement with (\ref{AA0}).

So far, the transformations considered have mapped $S$ back onto itself,
up to changed values of its parameters. But it is sometimes possible
to reduce more complex systems to the ones discussed here. For example,
the system $S$ with the parameters
\BEQ
D=1 \;\; , \;\; \delta=2\alpha +2\gamma
\EEQ
and $\gamma\neq 0$
gives through a SST the system
$\wit{S}$ with (see table~\ref{tab1})
\BEQ
\wit{D} = \frac{1}{3} ( 3+2\alpha+2\gamma) \;\; , \;\;
\wit{\alpha}=0 \;\; , \;\;
\wit{\gamma}=\frac{1}{3} (\alpha+\gamma) \;\; , \;\;
\wit{\delta}=2 \wit{\gamma} \;\; , \;\;
\wit{\nu} = 4\wit{\gamma}
\EEQ
Since the system $\wit{S}$ contains particle creation as well as particle
destruction terms, it no longer has a trivial (i.e. empty)
steady state. This steady
state can be found easily, since for a single-site state
\BEQ
B \left( \vekz{1-\rho}{\rho}\right) =
\left( \vekz{\frac{1}{3} - \frac{\gamma}{3\alpha+3\gamma} \rho}
{\frac{2}{3} + \frac{\gamma}{3\alpha+3\gamma} \rho} \right)
\EEQ
Since the steady state of $S$ is just $\bigotimes \left(\vekz{1}{0}\right)$,
we find that the steady state of $\wit{S}$ has a mean particle density
$\bar{\rho}=2/3$. The approach towards this steady state is exponential.

Another example is found when $\alpha+\gamma < \delta < 2\alpha+\delta$.
Then $S$ is similar to $\wit{S}$ with
\BEQ
\wit{D} = \frac{2}{3}(\delta-\alpha-\gamma)+1 \;\; , \;\;
\wit{\alpha} =  \frac{4}{3} (\delta-\alpha-\gamma) \;\; , \;\;
\wit{\gamma}=\wit{\delta} = \wit{\sigma}=
\frac{2}{3} (2\alpha+2\gamma-\delta) \;\; , \;\;
\wit{\beta} =\frac{1}{3} \delta
\EEQ
and the one-site state changes into
\BEQ
B \left( \vekz{1-\rho}{\rho} \right) =
\left( \vekz
{\frac{2}{3}+\frac{\delta-2\alpha-\gamma}{3(\delta-\alpha-\gamma)}\rho}
{ \frac{1}{3} - \frac{\delta-2\alpha-\gamma}{3(\delta-\alpha-\gamma)}\rho}
\right)
\EEQ
and we get a steady state particle density of $\bar{\rho}=1/3$. The
approach towards this steady state is exponential. The transformation
$\wit{S} \rar S$ is
a SST if $\delta > \frac{4}{3}\alpha +\gamma$. Conversely,
$S\rar\wit{S}$ is a SST
if $\delta < \frac{4}{3}\alpha +\gamma$.
Other examples with $\delta=0$ are given in \cite{Simon95}.

We now turn our attention to some systems with {\em biased} reaction-diffusion
processes. We take diffusion, coagulation and
annihilation into account. The rates are given in (\ref{Kopp}).
Using the unbiased case $\eta=0$ as a guide, we seek a SST
$S\rar\wit{S}$ such that
$\wit{\gamma}_{L,R}=0$. In fact, using $b_1=1$ as before and choosing
$b_2$ in order to get $\wit{\gamma}_R=0$, we find
\BEQ
\wit{\gamma}_L = \frac{ 4(\alpha+\gamma-1) \gamma\eta}{2\alpha+\gamma
+\eta(\gamma-2) } \;\; , \;\; \wit{\eta}=\eta \;\; , \;\; \wit{D} = D =1
\EEQ
For $\eta=0$, we recover the previous result. However, if we use the
free fermion condition $\alpha+\gamma=1$, then $\wit{\gamma}_L=0$ and
$\wit{\alpha}=1$. We then have
\BEQ
C_k(\{x\},t;\alpha,\gamma=1-\alpha;\eta;\rho) =
\left( \frac{2}{1+\alpha} \right)^k
C_k(\{x\},t;1,0;\eta;\frac{1}{2}(1+\alpha)\rho)
\EEQ
Considering the mean particle density only, this relation was also observed
in \cite{Priv95} for the special case of mapping the pure biased
coagulation problem ($\alpha=0$) onto the pure biased annihilation
problem ($\gamma=0$). Generally one finds that
for a homogeneous initial condition (\ref{IniS}) with $\rho_i \equiv \rho
=const$ the correlation function $C_k(\{x\},t;1,0;\eta;(1+\alpha)\rho/2)$
is independent of the bias $\eta $
\cite{Schue95a}. For an inhomogeneous initial state with $\rho_{x_0}=1$ and
$\rho_y =1/2$ for $y\neq x_0$ one finds for the
average density in an infinite system for large times \cite{Schue95a}
\BEQ
C_1(x,t) = <n_x>(t) = \frac{1}{\sqrt{4 \pi t}} + \frac{1}{\pi t^2}
 e^{-(x-x_0-\eta t)^2/2t}
\EEQ

Finally, we illustrate the transformation (\ref{BTra}) between a stochastic
and a non-stochastic system.
As an example, consider the Hamiltonian $H=\sum_i H_{i,i+1}$
\BEQ\label{XXH}
H_{i,i+1} = \left( \begin{array}{cccc}
A & 0 & 0 & - 2a \\
0 & A -1  & -D & 0 \\
0 & -D & A -1  & 0 \\
0 & 0 & 0 & A -2 \end{array} \right)
\EEQ
$H$ can be solved in terms of free fermions. We want the transformed
Hamiltonian $\wit{H}$ to describe a stochastic system, that is we require
that $\bra{s}\wit{H}=0$, see (\ref{LeftGS}). Writing
$B=\left(\matz{b_{11}}{b_{12}}{b_{21}}{b_{22}}\right)$,
the solution to this condition is
\BEQ
b_{21} = - b_{11} \;\; , \;\; A=2
\EEQ
We now take $b_{12}=b_{22}$. Let $\Gamma = ( b_{11}/b_{22} )^2 > 0$. Then the
positivity of the reaction rates in $\wit{H}$ requires that
$\Gamma= D/a$. The Hamiltonian then reads
\BEQ
\wit{H}_{i,i+1} = \left( \begin{array}{cccc}
2-2D & -1-D & -1-D & 0 \\
-1+D & 2+2D & 0 & D-1 \\
-1+D & 0 & 2+2D & D-1 \\
0 & -1-D & -1-D & 2-2D
\end{array} \right)
\EEQ
The off-diagonal elements of $\wit{H}$ are non-positive provided
$0 < D \leq 1$ and $0<a$. Under these conditions $B$ is non-singular and
$\wit{H}$ is the quantum Hamiltonian of a
stochastic system. We point out that this Hamiltonian is identical
to the quantum Hamiltonian for the $1D$ Glauber-Ising model \cite{Glauber}
\BEQ
H_{GI} = \sum_{i=1}^{L} \left( 1-\sig_{i}^{x} \right)
\left( 1 -\frac{D}{2} \left( \sig_{i}^{z}\sig_{i+1}^{z} +
\sig_{i-1}^z\sig_{i}^z \right) \right)
\EEQ
at temperature $T$ given by $D=\tanh{(2J/k_B T)}$.
In this way we obtain a new relation between non-zero temperature
Glauber dynamics and the $XXZ$ chain in the free fermion case.
On the other hand, for non-vanishing $T$ the Glauber-Ising model can
be transformed into an $XY$ free fermion chain \cite{Sigg77}. We shall return
to consequences of this observation and the reformulation of more
general stochastic
processes in terms of soluble free fermion systems elsewhere.

In summary, we have shown how to relate
different stochastic systems using similarity transformations. In several
examples, this technique proves to be useful to extend the scope of
integrable systems. The results are in agreement with the existing
experiments and include some previous observations of relations between
different systems as special cases. Going beyond similarity transformations
between stochastic systems, we have found a simple example how
to reformulate a stochastic system in terms of a non-stochastic
soluble free-fermion model in a novel way. The techniques developed here
can be used for a systematic study and classification
of stochastic interacting
particle systems in terms of integrable quantum chains.

\zeile{1}
\noindent{\bf Acknowledgements}
\zeile{1} \noindent
We thank K Krebs and H Simon for useful correspondence.
MH, EO and GMS were supported by the EC `Human Capital and Mobility'
programme.

\newpage
\appsection

Here we study the annihilation process for unstable
particles with an average life time $\gamma=\delta$.
The equations of motion for the average density
and density correlations read
\begin{eqnarray}\label{A1}
\frac{d}{dt} <n_x> & = &
    <n_{x+1}> + <n_{x-1}> - 2(1+\delta) <n_{x}> \nonumber \\
                   &   & - 2\alpha
    \left( <n_x n_{x+1}> + <n_{x-1} n_{x}>\right) \\
\label{A2}
\frac{d}{dt} <n_x n_{y}>  & = &  <n_{x+1} n_{y}> +  <n_{x-1} n_{y}> +
    <n_{x} n_{y+1}> + <n_{x} n_{y-1}> - 4(1+\delta) <n_{x} n_{y}> \nonumber \\
 &  & - 2 \alpha \left(<n_x n_{x+1} n_y> + <n_{x-1} n_{x} n_y> +
      <n_x n_y n_{y+1}> + <n_x n_{y-1} n_{y}> \right) \nonumber \\
 & & \qquad ~ \qquad \;\;\; ({\rm if~} |y-x| \geq 2) \\
\label{A3}
\frac{d}{dt} <n_x n_{x+1}>  & = &  <n_{x-1} n_{x+1}> +
    <n_{x} n_{x+2}>  - 2(1+ \alpha + 2\delta) <n_{x} n_{x+1}> \nonumber \\
 &  & - 2 \alpha \left( <n_{x-1} n_{x} n_{x+1}> + <n_x n_{x+1} n_{x+2}> \right)
\end{eqnarray}
and similar expressions for higher order correlators \cite{Schue95}. For a
$k$-point correlator there is always a coupling to $(k+1)$-point correlation
functions proportional to the annihilation rate $\alpha$.
First we show that the amplitude of a $k$-point density correlation function
$C_k(\{x\};t)=<n_{x_1} \dots n_{x_k}>$ decays
for large times $t$ at least with a factor
proportional to $\exp{(-2k\delta t)}$.

To see this, we recall that the spectrum of the quantum Hamiltonian $H$ is
exactly the same as of the XXZ quantum chain \cite{Alca94}. For the
calculation of the eigen{\em values}, it is thus sufficient to consider the
sectors with fixed number of particles $k$ separately. In the $k$-particle
sector, the eigenvalues of $H$ are
\BEQ \label{SpIE}
E_{\{i\}} = k 2 \delta + \sum_{i=1}^{N} 2 (1 - \cos q_i ) \geq k 2 \delta
\EEQ
and the values of the $q_i$ are
determined from the Bethe ansatz equations. On the other hand,
the $k$-point correlators $C_k$
can only take non-zero values when defined on states
which contain at least $k$ particles. Furthermore, $C_k$ depends through
the equations of motion directly only on $C_k$ and $C_{k+1}$ and in particular
it is independent of $C_0, C_1,\ldots, C_{k-1}$.
Thus, for large times $t$ we must have
\BEQ
C_k(\{x\};t) \sim e^{-\lambda t} \;\; ,
\;\; {\rm with}~ \lambda \geq k 2 \delta
\EEQ
because of the inequality (\ref{SpIE}) for the eigenvalues of $H$ in the
$k$-particle sector\footnote{Writing $C_k(\{x\};t)=
e^{-2 k \delta t}B_k(\{x\};t)$, one can further show with the Bethe ansatz
that generically $B_1(t)\rar O(1)$ and
$B_{k\geq 2}(t)\rar 0$ as $t\rar\infty$.}.

Now we define the explicitly time-dependent quantity
$\tilde{n}_x(t) = \exp{(2\delta t)} \, n_x$.
According to the considerations of the previous paragraph
$k$-point correlation functions of
this quantity are bounded from above by their
initial values $<\tilde{n}_{x_1} \dots \tilde{n}_{x_k}>(t=0) =
<n_{x_1} \dots n_{x_k}>(t=0) \leq 1$.
Rewriting Eqs. (\ref{A1}) - (\ref{A3})
and the corresponding equations for $k$-point correlators
in terms of averages for
$\tilde{n}_{x_i}$ reproduces equations of
the same form, but with effective time-dependent couplings $\tilde{\alpha}(t)
=\alpha\exp{(-2\delta t)}$ to $(k+1)$-point
correlators and with $\tilde{\delta}=0$:
\begin{eqnarray}\label{A4}
\frac{d}{dt} <\tilde{n}_x> & = &
<\tilde{n}_{x+1}> + <\tilde{n}_{x-1}> - 2 <\tilde{n}_{x}> \nonumber
                         \\
                   &   & - 2\alpha e^{-2\delta t}
\left(<\tilde{n}_x \tilde{n}_{x+1}> + <\tilde{n}_{x-1} \tilde{n}_{x}>\right) \\
\label{A5}
\frac{d}{dt} <\tilde{n}_x \tilde{n}_{y}>  & = &
<\tilde{n}_{x+1} \tilde{n}_{y}> +  <\tilde{n}_{x-1} \tilde{n}_{y}> +
    <\tilde{n}_{x} \tilde{n}_{y+1}> + <\tilde{n}_{x} \tilde{n}_{y-1}>
- 4 <\tilde{n}_{x} \tilde{n}_{y}> \nonumber \\
 &  & - 2 \alpha e^{-2\delta t}
\left( <\tilde{n}_x \tilde{n}_{x+1} \tilde{n}_y> +
<\tilde{n}_{x-1} \tilde{n}_{x} \tilde{n}_y> +
      <\tilde{n}_x \tilde{n}_y \tilde{n}_{y+1}> +
<\tilde{n}_x \tilde{n}_{y-1} \tilde{n}_{y}> \right) \nonumber \\
 & & \qquad ~ \qquad \;\;\; ( {\rm if~} |y-x| \geq 2) \\
\label{A6}
\frac{d}{dt} <\tilde{n}_x \tilde{n}_{x+1}>  & = &
<\tilde{n}_{x-1} \tilde{n}_{x+1}> +
    <\tilde{n}_{x} \tilde{n}_{x+2}> -
2(1+ \alpha) <\tilde{n}_{x} \tilde{n}_{x+1}> \nonumber \\
 &  & - 2 \alpha e^{-2\delta t}
\left( <\tilde{n}_{x-1} \tilde{n}_{x} \tilde{n}_{x+1}> +
<\tilde{n}_x \tilde{n}_{x+1} \tilde{n}_{x+2}> \right)
\end{eqnarray}
Since $<\tilde{n}_{x_1} \dots \tilde{n}_{x_k}>(t)\leq 1$
for all times $t$ and since
for long times $\alpha e^{-2\delta t}$ can be
neglected these equations effectively decouple and reduce to closed linear
differential-difference equations for $k$-point correlators which can be
solved with the Bethe ansatz \cite{Schue95}.

The equation for the one-point
function reduces to a lattice diffusion equation which is solved by
modified Bessel functions. Assuming a translationally invariant initial state
with
\newline $<n_x(t=0)>=<\tilde{n}_x(t=0)>=\rho_0$ one finds
for the average particle
density at time $t$
\begin{equation}
<n_x(t)> \approx \rho_0 e^{-2\delta t}
\end{equation}
This quantity depends on the initial density which is in contrast to the
diffusion limited annihilation of stable particles where the density decays
algebraically (for long times) and with an
amplitude independent of the initial density.

Defining $\tilde{C}(r,t)=<\tilde{n}_x\tilde{n}_{x+r}(t)>$ and using
(\ref{A4}) - (\ref{A6}) one finds for
the two-point function with translationally invariant initial conditions
the equations
\begin{eqnarray}
\frac{d}{dt} \tilde{C}(r,t) & = & 2\left( \tilde{C}(r+1,t) +\tilde{C}(r-1,t)
  - 2 \tilde{C}(r,t)\right) \;\;\; (r \geq 2) \nonumber \\
\frac{d}{dt} \tilde{C}(1,t) & = & 2\left( \tilde{C}(2,t)
  - (1+\alpha) \tilde{C}(r,t)\right)
\end{eqnarray}
This gives for the two-point density correlation
function $C(r,t)=<n_x n_{x+r}(t)>$ in an infinite system
\begin{equation}
C(r,t) \approx e^{-4(1+\delta)t} \sum_{y=1}^\infty \left[ a_y I_{r-y}(4t) +
b_y I_{r+y-1} (4t) \right]
\end{equation}
where $I_n$ is the modified Bessel function, $a_y$ are constants defined by
the initial distribution and $b_y = \mu a_y - (1-\mu^2) \sum_{k=1}^{y-1}
\mu^{y-1-k} a_k$ with $\mu=1-\alpha$.At first sight, we should expect for
large times that $C(r,t) \sim t^{-1/2} \exp(-4\delta t)$.
For the free fermion case $\alpha=1$, however,
a different result is found. Since $b_y = - a_{y-1}$ we get
\begin{equation}
C(r,t) \approx e^{-4(1+\delta)t} \sum_{y=1}^\infty a_y \left[I_{r-y}(4t) -
I_{r+y} (4t) \right]
\end{equation}
For $r^2 \ll t$ this correlator decays as
$C \sim t^{-3/2}\exp{(-4\delta t)}$ whereas
for $r^2 \sim t$ one has $C \sim t^{-1/2}\exp{(-4\delta t)}$.
The same effect is also seen for $\alpha=2$.

\end{document}